\documentclass[11pt,twoside]{article}

\usepackage{galev06}
\usepackage{epsf}
\usepackage{psfig}
\usepackage{lscape}

\begin{document}
\title{A Herschel survey of local galaxies activity (HERLOGA)}   
\author{Luigi Spinoglio}   
\affil{Istituto di Fisica dello Spazio Interplanetario - INAF, Via Fosso del Cavaliere 100, I-00133 Roma, Italy}    

\begin{abstract} 
The key project HERLOGA to be proposed for the Herschel mission is aimed to provide an homogeneous data set on local active galaxies that could be the reference for further studies on the interplay between star formation processes and accretion onto massive black holes in galactic nuclei. The selected sample will be the overlap between the 12$\mu$m sample and an all-sky hard-X ray selected sample of AGNs. The Herschel data will follow extensive observational work that is already available from radio to X-rays frequencies and will provide the baseline of zero-redshift objects for comparison with the high redshift Universe which will be revealed by the Herschel cosmological surveys.
\end{abstract}

\section{What we learned from ISO spectroscopy}

ISO opened a new window on the spectra of IR-bright, ultraluminous infrared galaxies and active galactic nuclei, followed in the last years by \textit{Spitzer}. The first mid-infrared spectra of active and starburst galaxies were obtained with ISO-SWS in Seyfert (Sturm et al. 2002) and starburst galaxies (Verma et al. 2003). \textit{Spitzer} spectroscopy followed up with more systematic studies of samples of Seyfert (e.g. Buchanan et al. 2006) and starburst galaxies. The MIR spectral range includes many fine-structure lines excited by the hard radiation produced by black hole accretion as well as those excited by stellar ionization (e.g. Spinoglio \& Malkan 1992) and MIR spectra can distinguish between the two processes, especially in obscured nuclei. However, while MIR spectroscopy can study the Narrow Line Regions (NLR) in obscured AGN and the effects of HII/starburst emission in actively star forming regions, FIR spectroscopy is the best way to study the influence of both the accretion process and stellar evolution on the geometry and physics of the densest, inner regions of galaxies. The geometry, and in particular the presence of the hypothetical torus around the nuclear engine, could be studied spectroscopically in the FIR, in a variety of molecular (high-J CO, OH, H$_2$O) and low excitation ionic transitions that trace the expected conditions of X-UV illuminated dusty tori (see e.g. Krolik \& Lepp 1989; for predictions of high-J CO lines).

\begin{figure}[h]
\centerline{\includegraphics[angle=0,width=6cm]{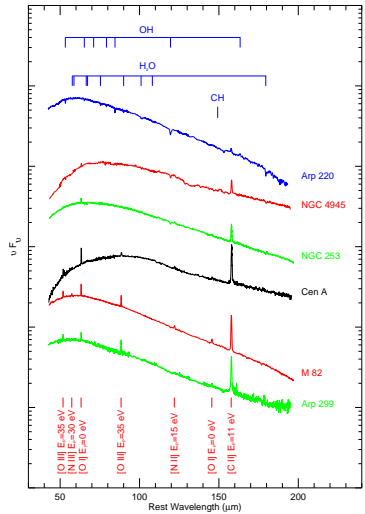}}
\caption{\footnotesize The full ISO-LWS spectra of six IR-bright galaxies.  The spectra have been shifted
and ordered vertically according to apparent excitation (Fischer
et~al. 1999) and are not in order of relative luminosity or
brightness.}\label{fig1}
\end{figure}

As can be seen in Fig.1, the ISO-LWS observations showed a dramatic progression in ionic fine-structure emission line and molecular/atomic absorption line characteristics from strong [OIII]52, 88$\mu$m and [NIII]57$\mu$m emission to detection of only faint [CII]157$\mu$m emission (Fischer et al.1999). The [CII]157$\mu$m line in 15 ULIRGs revealed an order of magnitude deficit compared to normal and starburst galaxies relative to the FIR continuum (Luhman et al 2003). LWS observations of Arp 220 show absorption in molecular lines of OH, H$_2$O, CH, NH, and NH$_3$, as well as in the [OI]63$\mu$m line and faint emission in the [CII]158$\mu$m line (Gonzalez-Alfonso et al 2004).  The complete 2-200$\mu$m spectrum of the prototype Seyfert 2 galaxy NGC1068 has been decomposed in the AGN and starburst components. Moreover, in the FIR it shows the 79, 119 and 163$\mu$m OH rotational lines in emission, not in absorption as in every other galaxy yet observed. Modeling the three FIR lines of OH suggests the gas lies in small (0.1pc) and dense clouds ($\sim 10^4 {\rm cm^{-3}}$)  most probably in the nuclear region (potentially a signature of the torus) (Spinoglio et al. 2005).

ISO FIR spectroscopic observations have been collected only for a few objects and a systematic work on a representative sample of galaxies has yet to be done. High sensitivity and high resolution FIR spectroscopy is needed to understand the effects of AGN and starbursts on the geometry and physics of galactic nuclear and circumnuclear regions. The interrelationship between star formation and accretion onto massive black holes and their connection with galaxy formation and evolution can be understood through detailed observations of active and starburst galaxies in the local universe. Because dust and active nuclei (AGNs) coexist even at very high redshifts (e.g., Priddey et al. 2003; Bertoldi et al. 2003), it is imperative to study nuclear activity in a way that minimizes the obscuring effects of dust. This implies that both the sample and the observations must be defined carefully to overcome the biases associated with dust, extreme star-formation rates, and nuclear activity. 

\section{The HERLOGA key project and its sample}

The HERLOGA (HERschel LOcal Galaxies Activity) key project is aimed at collecting 60-600$\mu$m spectroscopy and photometric imaging of local galaxies of a well studied statistical sample of local galaxies. Complementary \textit{Spitzer} low- and high resolution mid-IR spectra have been (Buchanan et al 2006) and are beeing collected (Spinoglio et al 2006, in prep.).
The spectroscopic observations with the Herschel spectrometers will be able to:
\begin{itemize}
\item Study the geometry and the physical conditions in nuclear and circumnuclear regions of galaxies. Detect the signature of the hypothetical torus, using OH, H$_2$O and high-J CO lines, good tracers of the expected conditions of X-UV illuminated dusty tori.
\item Study statistically the [CII]157$\mu$m line deficit and in general the suppression of the FIR fine-structure emission in starburst galaxies/ULIRGs. Is there any statistically significant far-IR spectroscopic difference between Seyfert and starburst galaxies?
\item Spatially - for nearby objects- and spectroscopically resolve the different excitation mechanisms (e.g. shocks, PDR, stellar and nonthermal photoionization).
\item Separate the non-thermal (AGN) and thermal (starburst) components in the energy budget of galaxies.
\end{itemize}

Complementary photometric observations with Herschel will:
\begin{itemize}
\item Determine how the obscuring material around AGN is distributed in terms of temperature, density and distance from the nucleus and whether the SED are consitent with torus models.
\item Measure how much star formation contributes to the FIR/submm luminosities of AGN, as a function of nuclear luminosity.
\item Determine the bolometric luminosities and the relative contributions of reprocessed radiation from dust and direct radiation from accretion and young stars. For the AGN, it is possible to derive the fraction of accretion power which is obscured in the local universe.
\end{itemize}
With accretion, star formation, obscuration and dust reprocessing all contributing to the spectral energy distributions of galactic nuclei, selection effects must be considered carefully. We consider that the optimum sample comes from a combination of two complementary \textit{all-sky} and \textit{flux-limited} samples, with orthogonal selection biases: the 12$\mu$m Galaxy Sample (hereafter 12MGS; Rush, Malkan \& Spinoglio 1993), and the HEAO-1 A2 sample of hard X-ray (2-10 keV) selected AGN (Piccinotti et~al. 1982; hereafter HX). The HX is selected purely on the basis of accretion radiation, it is unbiased with respect to infrared or host galaxy properties. The 12MGS is an IRAS-selected all-sky survey; it is the most representative sample of local infrared galaxies and is generally used to give the zero point to infrared cosmological studies of galaxies (e.g. Perez-Gonzalez et al 2005). It covers a wide range of nuclear activity - stellar and non-stellar, from HII/starbursts to Seyferts - allowing us to study the demographics, excitation mechanisms and observational signatures of these different classes.

Because of the large overlap between the two samples (3/4 of the HX belong to the 12MGS), they can be observed efficiently as a single programme. The combination of the HX and 12MGS sub-samples offers a key advantage for understanding the role of AGN in the IR galaxy population. While the HX sample allows us to investigate the FIR spectroscopic and photometric signatures of AGN as a function of accretion power, the 12MGS sample allows us to relate these properties to a coherent sample of nearby infrared galaxies. This in turn, forms the ideal reference sample with which to understand the large numbers of distant galaxies which will be discovered in Herschel's imaging surveys. As both the HX and 12MGS come from flux limited all-sky surveys, they provide statistically complete samples of the nearest and brightest objects over a wide range of luminosity (see Fig.2).

\begin{figure}[ht]
\centerline{\includegraphics[angle=-90,width=12cm]{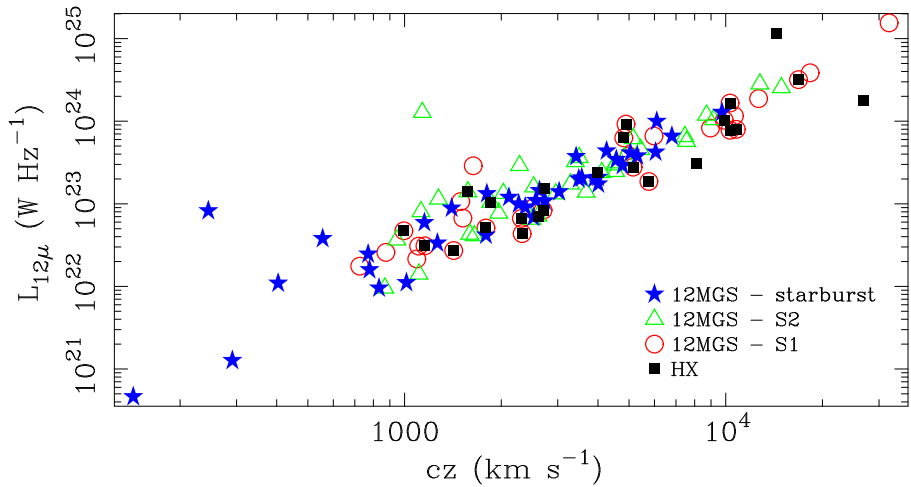}}
\caption{ \footnotesize Redshift and 12$\mu$m luminosity
distribution of the targets. S1 indicates Seyfert 1's and S2
Seyfert 2's. The targets are evenly distributed over more than 2.5
orders of magnitude in luminosity, and because they are drawn from
an all-sky flux-limited survey these are the closest (hence
brightest) targets at each luminosity.} \label{fig2}
\end{figure}

\end{document}